\documentclass[11pt,preprint]{aastex}

\usepackage{graphicx}

\newcommand{\ha}{\hbox{H$\alpha$}}
\newcommand{\hb}{\hbox{H$\beta$}}
\newcommand{\hei}{\hbox{He\,{\sc i}}}
\newcommand{\hg}{\hbox{H$\gamma$}}
\newcommand{\hd}{\hbox{H$\delta$}}

\newcommand{\hii}{\hbox{H\,{\sc ii}}}
\newcommand{\heii}{\hbox{He\,{\sc ii}}}

\newcommand{\nii}{\hbox{[N\,{\sc ii}]}}
\newcommand{\nv}{\hbox{[N\,{\sc v}]}}
\newcommand{\niii}{\hbox{N\,{\sc iii}}}
\newcommand{\caii}{\hbox{Ca\,{\sc ii}}}
\newcommand{\mgii}{\hbox{Mg\,{\sc i}+Mg\,{\sc h}}}

\newcommand{\xint} {\hbox{${\bf X}_{\rm INT}$}}
\newcommand{\xys} {\hbox{${\bf X}_{\rm YS}$}}
\newcommand{\xhii} {\hbox{${\bf X}_{\rm HII}$}}
\newcommand{\xold} {\hbox{${\bf X}_{\rm OLD}$}}
\newcommand{\xhy} {\hbox{${\bf X}_{\rm H+Y}$}}
\newcommand{\etal}{\hbox{et\thinspace al.\ }} 
\def\ni{\noindent}

\shorttitle{Spectroscopic study of BCGs: III}
\shortauthors{Kong et al.}

\begin{document}

\title{Spectroscopic study of blue compact galaxies \\
III. Empirical population synthesis}

\author{
X. KONG \altaffilmark{1,2},
S. Charlot\altaffilmark{1,3},
A. Weiss\altaffilmark{1},
F.Z. Cheng\altaffilmark{2}}

\altaffiltext{1}{Max Planck Institute for Astrophysics, Karl-Schwarzschild-Str. 
1, D-85741 Garching, Germany}
\altaffiltext{2}{Center for Astrophysics, University of Science and Technology 
of China, 230026, Hefei, P. R. China}
\altaffiltext{3}{Institut d'Astrophysique de Paris, CNRS, 98 bis Boulevard  
Arago, 75014 Paris, France}
\altaffiltext{4}{Xu KONG : xkong@mpa-garching.mpg.de; xkong@ustc.edu.cn}

\date{\centering Accepted for publication in A\&A: March 10, 2003}

\begin{abstract}
This is the third paper of a series dedicated to the study of the star 
formation rates, star formation histories, metallicities and dust
contents of a sample of blue compact galaxies (BCGs). We constrain the
stellar contents of 73 blue compact galaxies by analyzing their continuum 
spectra and the equivalent widths of strong stellar absorption features 
using a technique of empirical population synthesis based on a library
of observed star-cluster spectra. Our results indicate that blue 
compact galaxies are typically age-composite stellar systems; in 
addition to young stars, intermediate-age and old stars contribute 
significantly to the 5870 \AA\ continuum emission of most galaxies in
our sample. The stellar populations of blue compact galaxies also span
a variety of metallicities. The ongoing episodes of star formation started
typically less than a billion years ago. Some galaxies may be undergoing
their first global episode of star formation, while for most galaxies in
our sample, older stars are found to contribute up to half the optical 
emission. Our results suggest that BCGs are primarily old galaxies with 
discontinuous star formation histories.
These results are consistent with the results from 
deep imaging observations of the color-magnitude diagrams of a few 
nearby BCGs using HST and large ground-based telescopes.
The good quality of our population synthesis fits of BCG spectra allow
us to estimate the contamination of the \ha, \hb, \hg\ and \hd\ Balmer 
emission lines by stellar absorption. The absorption equivalent widths 
measured in the synthetic spectra range from typically 1.5\AA\ for \ha,
to 2--5\AA\ for \hb, \hg, and \hd. The implied accurate measurements of
emission-line intensities will be used in a later study to constrain 
the star formation rates and gas-phase chemical element abundances of 
blue compact galaxies.
\end{abstract}

\keywords{
galaxies: compact -- galaxies: evolution -- galaxies: stellar 
component -- galaxies: star clusters}

\section{Introduction}

The stellar populations of galaxies carry a record of their star 
forming and chemical histories, from the epoch of formation to the 
present. The global properties of galaxies are determined by the 
nature and evolution of their stellar components. Their studies thus 
provide a powerful tool to explore the physics of galaxy formation 
and evolution (\cite{kong00}; \cite{cid01}).  
For local group galaxies and some very nearby galaxies that can be 
resolved into individual stars with HST and 10-m class telescopes, 
the stellar population properties may be studied by means of direct
observations. However, for objects at larger distances, individual 
stars (except for some giants) are unresolved, even with 10-m class 
telescopes. The integrated light of such galaxies is expected to 
contain valuable information about their physical properties.

The most common approach used to interpret the integrated 
spectrophotometric properties of galaxies is {\em stellar population
synthesis}. There are two main types of population synthesis studies:
evolutionary population synthesis and empirical population synthesis
(EPS hereafter). Both types of studies require a complete library
of input (observed or model atmosphere) spectra of stars or star 
clusters (e.g., \cite{alvensleben00}). 

In the evolutionary population synthesis approach, pioneered by 
Tinsley (1967), the main adjustable parameters are the stellar initial
mass function (IMF), the star formation history and, in some cases, the
rate of chemical enrichment. Assumptions about the time evolution of 
these parameters allow one to compute the age-dependent distribution of
stars in the Hertzsprung-Russell diagram, from which the integrated 
spectral evolution of the stellar population can be obtained. In recent
years, a number of groups have developed evolutionary population 
synthesis models, which allow one to investigate the physical 
properties of observed galaxies (\cite{bruzual83}; Arimoto \& Yoshii 
1987; Buzzoni 1989; Bressan, Chiosi \& Fagotto 1994; Worthey 1994; 
Weiss, Peletier \& Matteucci 1995; Fioc \& Rocca-Volmerange 1997; 
Mas-Hesse \& Kunth 1999; Moeller, Fritze-v. Alvensleben \& Fricke 1999;
Leitherer et al. 1999; and Bruzual \& Charlot 1993, 2003). Such models
are convenient tools for studying the spectral {\em evolution} of 
galaxies, as they allow one to predict the past and future spectral 
appearance of galaxies observed at any given time. However, modern 
evolutionary population synthesis models still suffer from serious 
limitations (e.g., Charlot \etal\ 1996).

The empirical population synthesis approach, also known as `stellar 
population synthesis with a data base', was introduced by Faber 
(1972). In this technique, one reproduces the integrated spectrum of a
galaxy with a linear combination of spectra of individual stars or star
clusters of various types taken from a comprehensive library. The 
empirical population synthesis approach has been employed successfully
by several authors to interpret observed galaxy spectra (Faber 1972;
O'Connell 1976; Pickles 1985; Bica 1988; Pelat 1998; Boisson et al. 
2000; Cid Fernandes \etal 2001). An appealing property of this approach
is that the results do not hinge on a priori assumptions about stellar 
evolution, the histories of star formation and chemical enrichment, 
nor---in the case of a library of individual stellar spectra---the IMF.
In return, it does not allow one to predict the past and future 
spectral appearance of galaxies. In 1996, a workshop devoted to the 
comparison of various evolutionary population synthesis and empirical 
population synthesis codes showed broad agreement in general and a 
number of discrepancies in detail (Leitherer et al. 1996). 

In this paper, we use the empirical population synthesis approach
to constrain the stellar content and star formation histories of a 
sample of 73 blue compact galaxies (BCGs). Our approach builds on
earlier studies by Bica (1988) and Cid Fernandes \etal (2001), and we
adopt the library of observed star-cluster spectra assembled by Bica
\& Alloin (1986; see also Bica 1988). Our primary goal is to illustrate
the constraints that can obtained on the ages and metallicities of the
stellar populations of BCGs that cannot be resolved into individual 
stars, on the basis of their integrated spectra. In a forthcoming study,
we will use the results of this analysis to model the spectral 
evolution of BCGs using evolutionary population synthesis models. 
Another goal of the present work is to refine previous measurements of
the Balmer emission-line fluxes of the BCGs in our sample, through an
accurate modeling of the underlying stellar absorption spectrum. This
issue is critical to spectroscopic analyses of BCGs (e.g., 
\cite{thuan00}; \cite{olive01}). 

The paper is organized as follows. In Section 2, we briefly review the
principle of the empirical population synthesis approach. We describe 
the properties of the BCG spectral sample in Section 3. In Section 4, 
we estimate the stellar components in BCGs by fitting the observed 
equivalent widths and continuum colors with the empirical population 
synthesis approach. In Section 5 we discuss the age-metallicity 
degeneracy and the possible application of our results. Section 6
summarizes our main results.

\section{Empirical population synthesis model}

In a recent paper, Cid Fernandes \etal (2001) revisit the classical 
problem of synthesizing the spectral properties of a galaxy by using 
a `base' of star-cluster spectra, approaching it from a probabilistic 
perspective. Their work improves over previous EPS studies at both 
the formal and computational levels, and it represents an efficient 
tool for probing the stellar population mixture of galaxies. To provide
a quantitative description of the stellar components in the nuclear 
regions of BCGs, we will apply this EPS method to our BCG sample. It is
based on spectral-group templates built from star clusters of different
ages and metallicities, and on Bayes probabilistic theorem and the 
EPS Metropolis algorithm. In this section, we first describe the star 
cluster base. We then briefly review Bayes probabilistic formulation 
and the EPS Metropolis algorithm.

\subsection{Description of the star cluster base}
Empirical population synthesis by with a data base (a base of 
$n_{\star}$ spectra of stars or star clusters) is a classical tool 
designed to study the stellar population of galaxies. The recent EPS
algorithm of Cid Fernandes \etal (2001; see also Bica 1988) relies on a
data base of star-cluster spectra rather than on one of individual 
stellar spectra.  The star cluster base used by Cid Fernandes \etal 
(2001) was introduced by Schmidt \etal (1991). It consists of 12 
population groups, spanning five age bins --- 10 Gyr (representing
globular cluster-like populations), 1 Gyr, 100 Myr, 10 Myr and 
`HII'-type (corresponding to current star formation) --- and four 
metallicities --- 0.01, 0.1, 1 and 4 times Z$_\odot$ (see Table 1 of
Cid Fernandes \etal 2001). All metallicities are not available for
clusters of all ages. In particular, spectra of star clusters younger 
than 1 Gyr are only available for the metallicities Z$_\odot$ and 
4Z$_\odot$. Since blue compact galaxies are expected to have 
predominantly low abundances, the base of star-cluster spectra used by
Cid Fernandes \etal (2001) is not optimal for our analysis.

The base of 35 star-cluster spectra used by Bica (1988; see also Bica
\& Alloin 1986) is better appropriate for our purpose, as it contains 
spectra of young star clusters with metallicities down to 
$\log(Z/Z_\odot)=-0.5$ (see Table 1). Each component, corresponding to
a specific age and metallicity, is characterized by a set of six 
metallic features (\caii~K~$\lambda$3933, CN~$\lambda$4200, 
G~band~$\lambda$4301, \mgii~$\lambda$5175, \caii~$\lambda$8543,
\caii~$\lambda$8662) and three Balmer lines (\hd, \hg, and \hb), as
well as $n_C = 7$\ continuum fluxes at selected pivot wavelengths 
(3290, 3660, 4020, 4510, 6630, 7520 and 8700 \AA), which are normalized
at $\lambda$5870 \AA.  The library contains an HII region component 
corresponding to current star formation. This is represented by a pure
continuum based on the spectrum of 30 Dor, which is used at all 
metallicities (see the recent description by Schmitt \etal 1996).

Table 1 lists the ages and metallicities of each of the 35 components
of the Bica (1988) base of star-cluster spectra. The top line lists 
the ages of the components, while the rightmost column lists the
metallicities, with [Z/Z$_\odot$] = log(Z/Z$_\odot$).  

\begin{table}
\caption{Ages (in units of yr), metallicities and numbering convention
for the star clusters in the base.}
\vspace{1.0cm}
\centering
\begin{tabular}{ccccccccr}
\hline
\hline
HII& E7 & 5E7 & E8 &  5E8 &  E9 &  5E9 & E10&[Z/Z$_{\odot}$]\\
\hline
  35&    31&    27&    23&    19&    14&     8&     1&    0.6\\
  35&    32&    28&    24&    20&    15&     9&     2&    0.3\\
  35&    33&    29&    25&    21&    16&    10&     3&    0.0\\
  35&    34&    30&    26&    22&    17&    11&     4& $-$0.5\\
    &      &      &      &      &    18&    12&     5& $-$1.0\\
    &      &      &      &      &      &    13&     6& $-$1.5\\
    &      &      &      &      &      &      &     7& $-$2.0\\
\hline
\end{tabular}
\label{tab1}
\end{table}

\subsection{The synthesis algorithm}
The principle of EPS is to find the linear combination of a base of
spectra ($n_{\star}$ star clusters) that best reproduces a given set
of measured observables, such as the equivalent widths $W_j$ of $n_W$ 
conspicuous absorption features and the $n_C$ continuum fluxes $C_k$ in
an observed galaxy spectrum. Different synthesis algorithms have been 
developed to select the optimal combination of base spectra in the most 
efficient way. We use here the algorithm described by Cid Fernandes 
\etal (2001). Since this is relatively new, we briefly recall below its
probabilistic formulation and the main features of the algorithm.

The data ${\cal D}$\ we wish to model is composed of a set of 
$n_{obs} = n_W + n_C$ observables, as described above. The measurement
errors in these observables, collectively denoted by ${\bf \sigma}$, are
known from the observations. Given these, the problem of EPS is to 
estimate the population vector ${\bf X}$ (X$_i$, i=1,...,$n_{\star}$)
and the extinction $A_V$\ that `best' represents the data according to
a well defined probabilistic model, where X$_i$ denotes the 
fractional contribution of the $i$th base element to the total flux 
at the reference wavelength. The probability of a solution $({\bf 
X}, A_V)$ given the data ${\cal D}$ and the errors ${\bf \sigma}$, 
is given by Bayes theorem (\cite{smith85}):

\begin{equation}
P({\bf X}, A_V | {\cal D}, {\bf \sigma}, {\cal H}) = 
  \frac{P({\cal D} | {\bf X}, A_V,{\cal H})
  P({\bf X}, A_V | {\bf \sigma} , {\cal H})}
  {P({\cal D} | {\cal H})}.
\end{equation}
\ni ${\cal H}$ summarizes the set of assumptions on which the 
inference is to be made, 
$P({\cal D} | {\bf X}, A_V, {\bf \sigma}, {\cal H})$ is the 
likelihood,
$P({\cal D} | {\cal H})$ is the normalizing constant,
$P({\bf X}, A_V | {\cal H})$ is the joint {\it a priori} probability
distribution of ${\bf X}$ and $A_V$.

For a non-informative prior, the posterior probability
$P({\bf X}, A_V | {\cal D}, {\bf \sigma}, {\cal H})$ is simply 
proportional to the likelihood:

\begin{equation}
P({\bf X}, A_V | {\cal D}, {\bf \sigma}, {\cal H}) 
 \propto P({\cal D} | {\bf X}, A_V, {\bf \sigma}, {\cal H})
 \propto e^{-{\cal E}({\bf X}, A_V)} 
\end{equation}

\ni with ${\cal E}$ defined as half the value of $\chi^2$: 

\begin{eqnarray}
\label{eq:energy}
{\cal E}({\bf X}, A_V) 
  & = &
    \frac{1}{2} \chi^2({\bf X},A_V) 
    =
    \frac{1}{2}  \sum_{j=1}^{n_W}
    \left( \frac{ W_j^{obs} - W_j({\bf X}) }{\sigma(W_j)} \right)^2 
    \\ \nonumber
  & &
    + \frac{1}{2}  \sum_{k=1}^{n_C}
    \left( \frac{ C_k^{obs} - C_k({\bf X},A_V) }{\sigma(C_k)} 
\right)^2 \,.
\end{eqnarray}
Here $W_j^{obs}$ and $C_k^{obs}$ are the observed features and
$W_j({\bf X})$ and $C_k({\bf X},A_V)$ are the synthetic features.

This expression contains the full solution of the EPS problem, as 
embedded in it is not only the most probable model parameters but 
also their full probability distributions. In order to compute the 
individual posterior probabilities for each parameter, we use an
efficient parameter-space exploration method, known as the Metropolis 
algorithm (\cite{metropolis53}). The code preferentially visits 
regions of large probability, starting from an arbitrary point of the
parameter space. At each iteration $s$, we 
pick one of the $n_{\star}+1$ variables at random and change it by 
a uniform deviate ranging from $-\epsilon$ to $+\epsilon$, 
producing a new state $s+1$. Moves to states of smaller $\chi^2$ 
are always accepted, whilst changes to less likely states are 
accepted with probability exp[$-(\chi^2_{s+1} - \chi^2_s)$], thus 
avoiding trapping onto local minima. Moves towards non-physical 
regions ($X_i < 0$ or $ > 1$, $A_V <0$) were truncated. In this way, 
the probability distributions for the $X_i$ is given, and then the 
whole set ($X_i$, i=1,...,$n_{\star}$) is renormalized to unit sum.

The main output of the EPS approach is the {\it population 
vector} ${\bf X}$, whose $n_{\star}$ components carry the 
fractional contributions of each base element to the observed flux 
at the normalization wavelength 5870 \AA. This vector corresponds 
to the {\it mean} solution found from a $10^8$ steps 
likelihood-guided Metropolis walk through the parameter space. 
Owing to intrinsic errors in the observable parameters and some 
other uncertainties, more than one acceptable solution can 
represent the observation data. The final {\it mean} solution is 
given by the weighted ($e^{-\chi^2/2}$) average of all solutions 
within the observational errors. This {\it mean} solution is 
more reliable than the single optimal solution, and it provides a 
more representative result to the population synthesis problem.

\section{Description of the BCG sample}

Our sample is drawn from the atlas of optical spectra of 97 blue 
compact galaxies by Kong \& Cheng (2002a). The spectra were acquired
with the 2.16 m telescope at the XingLong Station 
of the Beijing Astronomical Observatory (BAO) in China.  A 300 line 
mm$^{-1}$ grating was used to achieve coverage in the wavelength 
region from 3580 to 7600 \AA\ with the dispersion is 4.8 
\AA\,pixel$^{-1}$.  The slit width was adjusted in between 
2$^{\prime\prime}$ and 3$^{\prime\prime}$ each night, depending on
seeing conditions.  A detailed decription of the 
observations, the sample selection, the data reduction and calibration
and the error analysis can be found in the first paper of this 
series (\cite{kong02a}). The average signal-to-noise ratio of the
spectra is $\sim$ 51 per pixel. The spectrophotometry is accurate to 
better than 10\% over small wavelength regions and to about 15\% or 
better on large scales.  

Kong \etal (2002b) measured several quantities in the spectra of 
the 97 BCGs in this sample, including the fluxes and equivalent widths
of emission lines, continuum fluxes, the 4000 \AA\ break and the 
equivalent widths of several absorption features. The galaxies were 
ordered into three classes based on their emssion-line properties: 13 
were classified as `non-emission-line galaxies' (non-ELG), 10 as 
`low-luminosity active galactic nuclei' (AGN) and 74 as `star-forming
galaxies' (SFG). We are primarily interested here in the star formation
rates, metallicities and star formation histories of BCGs. Therefore, 
we focus on the subsample of 74 star-forming galaxies (SFGs). We 
exclude I Zw 207 from this subsample because the absence of a blue 
spectrum prevents the measurements of most absorption features in
this galaxy. Our final sample therefore consists of 73 star-forming, 
blue compact galaxies.

\section{Empirical population synthesis results}

We use the EPS approach of Cid Fernandes \etal (2001; Section 2.2 
above) and the library of star-cluster spectra of Bica (1988; Section
2.1 above) to interpret the spectra of the 73 star-forming BCGs in our
sample.

\subsection{Input data}

We use the following observable quantities to constrain the stellar 
components in the nucleus regions of BCGs: the observed absorption 
equivalent widths ($W_j^{obs}$; 
$W_j^{obs} \ge 1.0$ \AA) of \caii\ K $\lambda$3933, \hd\ $\lambda$4102,
CN $\lambda$4200, G band $\lambda$4301, \hg\ $\lambda$4340, and 
\mgii\ $\lambda$5176 and the continuum fluxes (normalized at 5870 \AA)
at 3660, 4020, 4510, 6630, and 7520 \AA\ (in practice, we use the \hd\
and \hg\ absorption equivalent widths only for galaxies with negligible
emission at these wavelengths, representing roughly half of the sample).
The absorption equivalent widths and continuum fluxes were measured 
by Kong \etal (2002b) according to the procedure outlined in Bica 
(1988) and Cid Fernandes \etal (2001; see Kong \etal 2002b for detail).
In some spectra, the continuum flux at 5870 \AA\ may be buried 
underneath the \hei\, $\lambda$5876 \AA\ emission line. In such cases, 
adjacent wavelength regions were used to estimate the continuum level.
The resulting absorption equivalent widths are listed in Table 6 of Kong
\etal (2002b), and the continuum fluxes in Table 4 of Kong \etal 
(2002b). The errors in the absorption equivalent widths and 
continuum fluxes must also be included as input parameters. These are
set to 10\% for those absorption bands with $W_j^{obs} > 5$\AA, 
20\% for those absorption bands with $1.0 <W_j^{obs} < 5$ \AA, and 
10\% for continuum colors. These errors are consistent with the quality
of the spectra (Kong \& Cheng 2002a).

\subsection{Computation}

To select the linear combination of the base cluster spectra that
best represents an observed galaxy spectrum, we set the EPS algorithm
to sample $N_s = 10^8$ states of the whole age versus [Z/Z$_\odot$] 
parameter space, with the `visitation parameter' $\epsilon$ set to 0.05
for the $X_i$'s ({\bf $\Delta X_i$ = 0.05}) and 0.01 for $A_V$. As 
described in Section 2.2, we obtain as output a 35-dimension, {\em 
mean} population vector ${\bf X}$ containing the expected values of the 
fractional contribution of each component to the total light at the 
normalization wavelength $\lambda$5870 and $A_V$. Table~2 lists, as
examples, the results for two galaxies in our sample.

\begin{table*}
\caption{Empirical population synthesis results of percentage contributions
from stars of all ages and metallicities to $F_{\lambda5870}$, for two sample
galaxies. The `evolutionary paths' favored by the analysis are highlighted 
(see text).}
\vspace{1.0cm}
\tiny
\centering
\begin{tabular}{rrrrrrrrrrrrrrrrrrr}
\hline
\hline
\multicolumn{8}{l}{a) Mrk 385 \hspace{1.5cm}     Age (yr)  }&[Z/Z$_\odot$]&&
\multicolumn{8}{l}{b) I Zw 97 \hspace{1.5cm}     Age (yr)  }&[Z/Z$_\odot$]\\
\cline{1-8}\cline{11-18}
HII& E7 & 5E7 & E8 &  5E8 &  E9 &  5E9 & E10 & &&
HII& E7 & 5E7 & E8 &  5E8 &  E9 &  5E9 & E10 & \\
\cline{1-9}\cline{11-19}
   &2.0&1.8&2.1&  .3& .2& .2& .5& 0.6&&    & .3&1.6&2.6&1.7& .3& .2&  .1& 0.6\\
   &2.1&2.3&2.9&  .5& .3& .3& .6& 0.3&&    & .4&1.9&3.3&3.0& .6& .3&  .1& 0.3\\
   &1.5&3.3&3.9&  .8& .5& .5& .8& 0.0&& 
{\bf8.0}&{\bf.8}&{\bf5.9}&{\bf9.8}&{\bf5.9}&1.7& .6& .2& 0.0\\
{\bf8.8}&{\bf4.0}&{\bf6.0}&{\bf6.9}&{\bf11.7}&1.7&1.5&1.8
&-0.5&&    & .5&1.9&3.3&1.8&2.8&1.2& .7&-0.5\\
   &   &   &   &    &{\bf4.5}&3.4&3.4&-1.0&&    &   &   &   &   &{\bf4.7}&1.9& 1.5&-1.0\\
   &   &   &   &    &   &{\bf5.5}&5.4&-1.5&&    &   &   &   &   &   &{\bf6.9}&7.0&-1.5\\
   &   &   &   &    &   &   &{\bf8.1}&-2.0&&    &   &   &   &   &   &   &{\bf16.7}&-2.0\\
\hline
\end{tabular}
\end{table*}

The results for the 73 BCGs in our sample indicate that, in all cases,
a single metallicity is favored for all four youngest stellar 
components (corresponding to ages less than $5\times10^8{\rm yr}$). 
The metallicity of young stars is [Z/Z$_\odot$]=--0.5 for 55 galaxies
in the sample and [Z/Z$_\odot$]=0.0 for the remaining 18 galaxies.
For stars older than $5\times10^8{\rm yr}$, the dominant metallicity
anticorrelates with age, in the sense that the metallicity of the
stellar component contributing most to the integrated light increases
with decreasing age from $1\times10^{10}{\rm yr}$ to $1\times10^{9}{\rm
yr}$ (see the examples in Table~2). We note that the dominant 
metallicity of stars older than $5\times10^8{\rm yr}$ is always
found to be less than or equal to that of younger stars. It is not 
surprising that a dominant metallicity be favored for stars of any 
given age in a galaxy. The nuclei of BCGs correspond to small volumes, 
where star-forming gas is expected to be chemically homogeneous at any
time. It is worth noting that our results differ from those of previous
EPS analyses, such as those performed by Bica (1988) and Schmitt \etal 
(1996), in that we determine the evolutionary paths of galaxies in the 
age-metallicity plane from a full maximum-likelihood analysis. In most 
previous studies, the star-formation and chemical-enrichment histories 
of galaxies were selected from a limited number of a priori 
evolutionary paths.

It is of interest to exploit these results and compute simple
`evolutionary paths' for the galaxies in our sample. As mentioned 
above, for each galaxy, the EPS analysis favors a dominant
metallicity for stars of any given age, such that young stars 
have typically a higher metallicity than older stars. At any age, 
however, stars of any metallicity are assigned non-zero weights
by the EPS algorithm. To represent the star-formation and 
metal-enrichment histories of BCGs in a schematic way, we adopt
for each galaxy the path favored by the EPS analysis in the 
age-metallicity plane. For consistency, we readjust the various 
proportions of stars of different ages along this path by rerunning the
EPS algorithm after setting the weights of all stars outside the path
to zero. Since, for a given galaxy, the age-metallicity space to be 
explored then reduces to a single dimension, we refine the analysis by
lowering the `visitation parameter' $\epsilon$ to 0.005 ({\bf $\Delta
X_i$ = 0.005}) while still sampling $N_s = 10^8$ states (we keep 
$\epsilon=0.01$ for $A_V$).

\begin{figure}
\centering
\includegraphics[angle=-90,width=\textwidth]{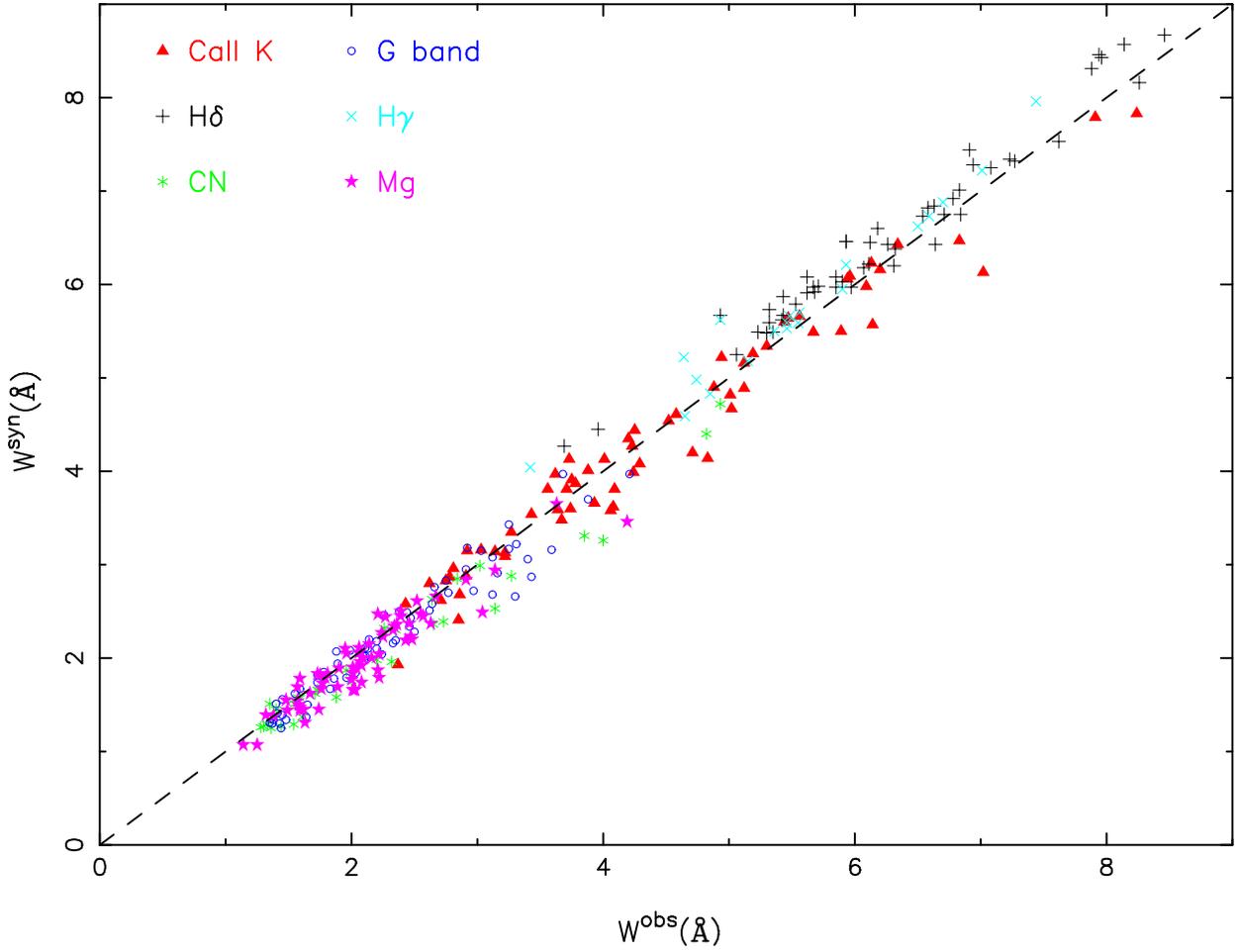}
\caption{
Synthetic equivalent width $W^{syn}$ against observed equivalent 
width $W^{obs}$ for all $W_j^{obs} \ge 1.0$ \AA\ absorption features
used to constrain the star-formation and chemical-enrichment histories
of 73 BCGs (see Tables 3 and 4). The dash line is the identity line. 
Different symbols denote different absorption features, as indicated.}
\label{eqw-syn-obs}
\end{figure}

\begin{table}
\caption{Empirical population synthesis results of the galaxies with 
asymptotic (young-star) metallicities [Z/Z$_\odot$]=$-$0.5: percentage 
contribution to $F_{\lambda5870}$.}
\scriptsize
\begin{tabular}{lrrrrrrrrrrrrrrrrrr}
\hline
\hline
\multicolumn{1}{l}{[Z/Z$_\odot$]}&
\multicolumn{5}{c}{[Z/Z$_\odot$]=$-$0.5}&&$-$1.0&&$-$1.5&&$-$2.0&&&&\multicolumn{4}{c}
{EW$_{abs}$(\AA)}\\
\cline{2-6}\cline{8-8}\cline{10-10}\cline{12-12}\cline{16-19}
Age(yr)&HII&E7& 5E7 & E8 &  5E8 &&  E9 &&  5E9 && E10&\\
\cline{9-12}
\hline
Name     &   35&   34&   30&   26&   22&&   18&&   13&&    7&&$A_V$&&\hd &\hg\ &\hb\ & \ha \\
\hline
iiizw12  & 18.2& 10.0& 16.2& 10.5& 10.9&& 18.9&&  9.6&&  5.6&& .07&& 4.62& 3.50& 4.09& 1.51\\
haro15   & 23.3&  5.7& 19.8& 20.8& 13.5&&  9.8&&  4.0&&  3.2&& .04&& 4.59& 3.27& 3.75& 1.51\\
iiizw33  & 21.3&  5.4& 10.7&  8.2& 18.5&& 16.5&& 11.3&&  8.1&& .08&& 3.78& 2.44& 3.56& 1.38\\
iiizw43  &  1.0&  1.6&  4.6&  6.1& 21.4&& 43.0&& 16.4&&  6.0&& .86&& 4.24& 3.45& 4.33& 1.25\\
iizw23   & 14.1&  6.0& 14.4& 10.2&  8.4&& 18.0&& 19.6&&  9.3&& .27&& 4.76& 3.17& 4.26& 1.89\\
iizw28   & 28.1&  6.3& 10.6&  7.8& 19.8&& 14.4&&  7.6&&  5.5&& .06&& 3.41& 2.19& 3.09& 1.03\\
iizw33   & 27.3&  3.5&  5.1&  3.0&  6.4&& 24.0&& 22.6&&  8.1&& .09&& 3.01& 1.88& 2.94& 1.59\\
iizw40   & 52.4&  1.8& 11.9& 13.1& 13.5&&  4.8&&  1.4&&  1.2&& .01&& 2.57& 2.11& 2.23&  .66\\
mrk5     & 12.1&  2.3&  7.2&  9.2& 48.6&& 14.6&&  3.1&&  2.9&& .02&& 3.57& 3.20& 4.43&  .28\\
viizw153 &  9.2&  5.1& 14.8& 11.3& 24.1&& 23.6&&  7.3&&  4.6&& .05&& 4.70& 3.81& 4.36& 1.53\\
viizw156 &  2.6&  2.8& 16.3& 15.8& 17.5&& 16.4&& 17.8&& 10.9&& .17&& 5.53& 4.02& 4.55& 1.71\\
haro1    & 21.7&  6.8& 14.8& 18.6& 15.2&&  9.8&&  6.1&&  6.9&& .09&& 4.37& 3.12& 4.76& 1.82\\
mrk385   &  9.0&  6.7& 16.4& 14.3& 22.6&& 13.3&&  9.2&&  8.6&& .10&& 5.03& 2.48& 4.60& 1.39\\
mrk390   & 25.0&  7.7& 14.7& 10.3&  8.6&& 18.1&& 10.4&&  5.4&& .11&& 4.08& 2.84& 3.81& 1.93\\
mrk105   &  8.5&  9.5&  8.5&  8.8& 12.9&& 25.0&& 16.2&& 10.6&& .41&& 4.69& 3.34& 4.25& 1.50\\
izw18    & 47.8&  1.1&  6.9& 10.5& 26.9&&  4.5&&  1.2&&  1.1&& .01&& 2.25& 1.87& 2.31&  .29\\
mrk402   & 23.5&  7.1& 12.9&  7.7& 14.7&& 24.4&&  6.3&&  3.4&& .04&& 3.80& 3.10& 3.49& 1.63\\
haro22   & 27.0&  3.9& 20.1& 15.5&  9.0&& 11.0&&  7.7&&  5.9&& .06&& 4.29& 3.35& 3.47& 1.55\\
haro23   & 28.4&  3.2& 12.4&  9.5& 19.6&& 19.0&&  4.7&&  3.3&& .03&& 3.39& 2.64& 3.35& 1.01\\
haro2    & 32.6&  5.5& 15.0& 13.5& 14.2&& 13.0&&  3.6&&  2.6&& .03&& 3.63& 2.77& 3.45& 1.64\\
mrk148   & 16.6&  4.3& 12.5&  9.2&  8.7&& 10.1&& 21.0&& 17.6&& .84&& 4.39& 2.57& 3.98& 1.69\\
haro25   & 26.5&  7.4& 14.5& 15.5& 10.4&&  8.4&&  7.9&&  9.5&& .10&& 4.11& 3.27& 3.84& 1.30\\
mrk1267  & 20.6&  4.3& 15.8& 16.3& 26.0&&  9.9&&  3.8&&  3.3&& .03&& 4.22& 3.08& 3.96&  .89\\
haro4    & 45.6&  3.6& 18.9& 14.5&  8.0&&  5.6&&  2.1&&  1.7&& .02&& 3.34& 2.66& 2.93& 1.00\\
mrk169   & 16.3&  9.2& 10.2&  6.6&  7.7&& 14.8&& 22.9&& 12.2&& .61&& 4.43& 3.07& 4.59& 1.88\\
haro27   & 13.9&  4.1&  6.8&  4.5& 14.7&& 35.8&& 13.2&&  7.0&& .07&& 3.66& 2.67& 3.23& 1.72\\
mrk201   & 24.5&  4.2& 12.8& 15.1& 11.6&& 11.6&&  9.4&& 10.7&& .12&& 3.96& 3.11& 4.02& 1.61\\
haro28   &  4.2&  3.7&  7.6&  5.4& 12.5&& 31.6&& 25.1&&  9.8&& .09&& 4.69& 2.03& 3.94& 1.44\\
haro8    & 11.7&  2.0&  8.3&  8.3& 30.6&& 36.2&&  1.8&&  1.2&& .01&& 3.63& 3.14& 3.91&  .80\\
haro29   & 37.4&  2.0& 12.4& 11.9& 25.3&&  7.7&&  1.8&&  1.4&& .01&& 3.04& 2.49& 3.12&  .78\\
mrk215   & 19.9&  8.7& 11.2&  8.5& 15.0&& 14.8&& 12.3&&  9.7&& .10&& 4.09& 3.11& 4.39& 1.67\\
haro32   & 36.9&  5.2& 12.9&  9.2&  8.2&& 16.2&&  6.9&&  4.4&& .05&& 3.23& 2.23& 2.66& 1.28\\
haro33   & 21.3&  2.5& 17.8& 16.3& 23.9&& 12.1&&  3.4&&  2.7&& .03&& 4.23& 3.20& 3.18& 1.24\\
haro36   &  7.8&  2.8&  8.4&  6.7& 43.3&& 23.6&&  4.2&&  3.2&& .02&& 3.83& 1.54& 3.59&  .52\\
haro35   & 19.7&  4.4& 17.2& 17.8& 14.6&& 14.5&&  6.3&&  5.4&& .05&& 4.50& 3.66& 4.54& 1.27\\
haro37   & 25.7&  6.9& 16.5& 12.9& 14.8&& 12.1&&  6.4&&  4.7&& .05&& 4.12& 3.23& 3.39& 1.30\\
mrk57    &   .6&  1.0&  3.1&  3.7& 29.0&& 37.5&& 20.0&&  5.2&& .17&& 4.02& 2.60& 3.77& 1.09\\
mrk235   & 15.2&  4.4&  7.4&  5.8& 10.4&& 20.2&& 23.0&& 13.6&& .14&& 3.98& 2.82& 3.40& 1.53\\
haro38   & 12.4&  2.8& 14.2& 11.4& 33.5&& 18.0&&  4.3&&  3.4&& .03&& 4.25& 2.88& 3.31& 1.19\\
mrk275   & 20.0&  3.9& 12.2&  7.0& 19.7&& 23.4&&  8.7&&  5.1&& .05&& 3.77& 2.60& 3.42& 1.25\\
haro42   & 23.3&  3.1&  9.8&  7.9&  9.7&& 22.6&& 14.3&&  9.3&& .07&& 3.58& 2.73& 3.22& 1.32\\
haro43   & 10.0&  3.8& 17.0& 12.1& 29.3&& 18.8&&  5.2&&  3.8&& .03&& 4.69& 3.30& 4.23& 1.32\\
haro44   & 25.9&  4.4& 20.8& 15.7& 18.5&&  9.6&&  3.0&&  2.1&& .02&& 4.28& 3.65& 3.54& 1.11\\
iizw70   & 38.2&  4.1& 13.1& 10.5& 19.4&&  9.9&&  2.8&&  2.1&& .02&& 3.11& 2.67& 3.02&  .84\\
izw117   & 22.3&  7.0& 13.1& 11.2&  8.8&& 11.4&& 14.0&& 12.1&& .56&& 4.18& 3.47& 3.76& 1.43\\
izw123   & 26.6&  3.9& 12.2& 12.7& 27.2&&  9.7&&  4.0&&  3.7&& .03&& 3.58& 1.58& 3.82& 1.10\\
mrk297   & 31.6&  3.4& 12.6& 14.9& 16.0&&  9.1&&  5.7&&  6.8&& .06&& 3.52& 2.51& 3.63& 1.51\\
izw159   & 22.0&  6.8& 11.7& 10.3& 23.1&& 17.4&&  5.2&&  3.7&& .04&& 3.81& 2.98& 3.60& 1.27\\
izw166   & 25.1&  7.6& 15.6& 11.7&  8.8&&  8.6&& 11.7&& 10.8&& .27&& 4.21& 2.23& 4.03& 1.85\\
mrk893   & 15.7&  6.2& 10.4&  7.6& 12.9&& 20.3&& 16.5&& 10.5&& .16&& 4.19& 2.83& 3.51& 1.98\\
izw191   &  5.0&  8.8&  7.5&  7.1& 14.6&& 33.6&& 15.3&&  8.0&& .19&& 4.71& 3.88& 3.92& 1.49\\
ivzw93   & 22.5&  5.0& 12.6&  7.7& 12.9&& 25.3&&  9.4&&  4.7&& .05&& 3.79& 2.56& 3.51& 1.82\\
mrk314   & 22.6&  4.1& 14.9& 10.1&  9.7&& 11.6&& 14.7&& 12.2&& .16&& 4.13& 2.22& 2.78& 1.57\\
ivzw149  & 16.0&  4.9& 16.6& 14.8& 23.7&& 16.0&&  4.5&&  3.5&& .03&& 4.49& 3.40& 4.17& 1.44\\
zw2335   & 16.5&  4.9& 11.1&  9.2& 11.3&& 13.4&& 19.3&& 14.4&& .34&& 4.27& 3.26& 3.92& 1.58\\
\hline
\end{tabular}
\end{table}

\begin{table}
\caption{Empirical population synthesis results of the galaxies with asymptotic 
(young-star)
 metallicities [Z/Z$_\odot$]= 0.0: percentage contribution to $F_{\lambda5870}
$.}
\scriptsize
\centering
\begin{tabular}{lrrrrrrrrrrrrrrrrrrr}
\hline
\hline
\multicolumn{1}{l}{[Z/Z$_\odot$]}&
\multicolumn{5}{c}{[Z/Z$_\odot$]= 0.0}&&$-$1.0&&$-$1.5&&$-$2.0&&&&\multicolumn{
4}{c}
{EW$_{abs}$(\AA)}\\
\cline{2-6}\cline{8-8}\cline{10-10}\cline{12-12}\cline{16-19}
Age(yr)&HII&E7& 5E7 & E8 &  5E8 &&  E9 &&  5E9 && E10&\\
\cline{9-12}
\hline
Name     &   35&   33&   29&   25&   21&&   18&&   13&&    7&&$A_V$&&\hd &\hg\ &\hb\ & \ha \\
\hline
vzw155   &  9.6&  6.2&  9.2&  7.7& 20.2&& 15.7&& 17.9&& 13.5&& .24&& 4.45& 2.68& 3.78& 1.61\\
iiizw42  & 16.4&  7.2& 12.7&  8.4& 15.1&& 11.6&& 14.5&& 14.1&& .54&& 4.35& 2.57& 3.80& 1.44\\
zw0855   & 11.5&  2.5&  7.5&  9.0& 49.7&& 15.0&&  2.4&&  2.5&& .03&& 3.59& 3.06& 3.50& -.07\\
iizw44   &  6.7&  5.6&  7.2&  5.5& 20.0&& 15.7&& 23.8&& 15.6&& .21&& 4.49& 3.50& 4.72& 1.43\\
mrk213   &  5.9&  5.1&  6.3&  5.8& 20.4&& 18.4&& 22.2&& 15.8&& .46&& 4.42& 3.46& 4.67& 1.55\\
iizw67   &  4.9&  4.2&  4.0&  3.6&  9.4&& 20.7&& 33.4&& 19.8&& .39&& 4.51& 2.67& 5.31& 1.70\\
mrk241   &  7.3&  4.1& 10.5&  7.8& 18.5&& 14.7&& 17.6&& 19.5&& .32&& 4.58& 2.79& 4.08& 1.26\\
izw53    & 17.7&  9.3&  7.4&  4.6& 26.7&& 12.1&& 13.3&&  8.9&& .24&& 3.77& 3.23& 3.64& 1.00\\
izw56    &  9.3&  7.7&  9.1&  5.6& 10.5&& 13.1&& 24.3&& 20.5&& .82&& 4.67& 2.74& 3.82& 1.87\\
iizw71   &  1.6&  2.8&  7.0&  5.9& 53.7&& 13.7&&  9.4&&  5.8&& .13&& 4.11& 2.23& 3.48&  .34\\
izw97    &  7.1&  2.5& 14.2& 16.0& 31.8&& 10.5&&  7.4&& 10.4&& .06&& 4.75& 3.38& 4.53& 1.20\\
izw101   & 11.7&  4.9&  8.5&  5.7& 17.9&& 15.1&& 19.0&& 17.1&& .45&& 4.16& 2.86& 3.56& 1.50\\
mrk303   &  9.1&  6.5&  4.1&  2.8& 14.4&& 17.6&& 30.5&& 14.8&& .23&& 4.23& 2.99& 4.02& 1.87\\
zw2220   & 22.4& 11.4& 15.8& 10.9&  4.7&&  8.3&& 13.8&& 12.7&& .45&& 4.56& 3.38& 3.97& 1.54\\
ivzw142  &  9.1&  9.7&  4.0&  3.2& 24.9&& 19.8&& 19.7&&  9.7&& .46&& 4.08& 3.35& 3.41& 1.32\\
\hline
\end{tabular}
\end{table}

\subsection{Stellar populations}

The results of our analysis are summarized in Tables 3 and 4 for the
galaxies with asymptotic (young-star) metallicities [Z/Z$_\odot$]=--0.5
and 0.0, respectively. For each galaxy, we report the percentage contributions
from each age-metallicity component to the integrated flux at $\lambda = 
5870$ \AA. The first and second lines of each table indicate the 
metallicities and the ages of the different stellar components. For 
reference, the third line indicates the indices of the stellar 
components in the cluster data base of Table 1. For each galaxy, we also list
the inferred $V$-band attenuation $A_V$ and the absorption equivalent widths
of \hg, \hd, \hb\ and \ha, as measured from the synthetic stellar population
spectra. Figure \ref{eqw-syn-obs} shows the equivalent widths ($W_j^{syn}$) 
of all the stellar absorption features used to constrain the fits in the 
model spectra against those in the observed spectra ($W_j^{obs}$, for 
$W_j^{obs} \ge 1.0$ \AA), for all 73 BCGs in our sample. Clearly, the feature
strengths in the model spectra are in very good agreement with those in the 
observed spectra.

A first noticeable result in Tables 3 and 4 is that all BCGs show an 
underlying component of stars older than $1\times10^{9}{\rm yr}$. The 
fractional contribution of this component to the total light at 
$\lambda$5870 \AA\ exceeds 15\% for most galaxies, except for some 
low-luminosity BCGs, such as I Zw 18, II Zw 40, which have
very strong emission line spectra, and have a marginally detected old 
component.  We note that the spectra used here sample the inner
regions of the galaxies. The contribution from old and intermediate-age
stars to the integrated light could be even larger in the extended 
(off-nuclear) regions of galaxies. The presence of significant populations
of old and intermediate-age stars indicates that blue compact galaxies have
experienced substantial episodes of star formation in the past. This 
supports the results from deep imaging observations of the color-magnitude
diagrams of a few nearby BCGs that these are old galaxies (see section
5.2 below). As expected, stars younger than $5\times10^8{\rm yr}$ tend to 
dominate the emission at 5870 \AA, consistent with the observational character
of BCGs, i.e. blue colors and strong emission lines.

Another interesting result of Tables 3 and 4 is that blue compact galaxies
present a variety of star formation histories. In II Zw 67 and I Zw 56, for
example, stars with age 10 Gyr contribute as much as 20 per cent of the flux
at $\lambda$5870 \AA. In contrast, in III Zw 43 and Mrk 57, old stars do not 
contribute significantly to the flux at $\lambda$5870 \AA, that is produced
in half by intermediate-age stars. The galaxies I Zw 18 and II Zw 40
differ from these cases in that their emission at $\lambda$5870 \AA\ is
accounted almost entirely by young stars. The star formation history of BCGs,
therefore, appears to vary significantly on a case by case basis.

\begin{figure*}
\centering
\includegraphics[angle=-90,width=0.95\textwidth]{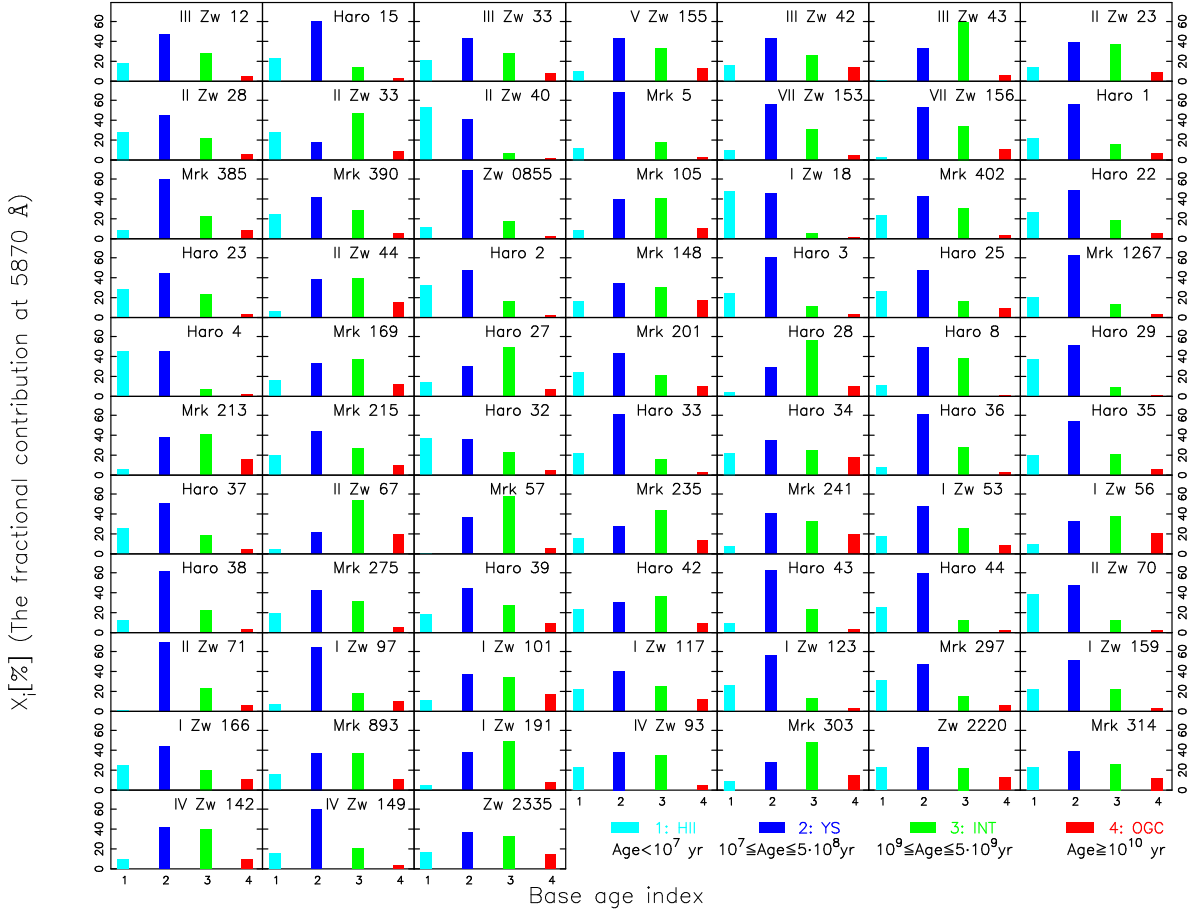}
\caption{
Star formation histories of 73 BCGs. In each panel, the numbers on the 
horizontal axis represent the stellar age groups, as indicated in the 
bottom right of the figure. The vertical axis shows the percentage 
contributions of stars in the various age groups to the integrated 
flux at 5870\AA.}
\label{fig-age}
\end{figure*}

For simplicity, in what follows we arrange stars into four age bins:
{\bf OLD} stars with age $1\times10^{10}{\rm yr}$ (proportion \xold); {\bf 
INT}ermediate-age stars with ages between $1\times10^{9}{\rm yr}$ and 
$5\times10^{9}{\rm yr}$ (proportion \xint); {\bf Y}oung {\bf S}tars with 
ages between $10^{7}{\rm yr}$ and $5\times10^{8}{\rm yr}$ (proportion 
\xys); and newly-born stars in \hii\ regions (proportion \xhii). Hence 
\xold + \xint + \xys + \xhii = 1. 
Figure \ref{fig-age} illustrates graphically the
variety of star formation histories reported in Tables~3 and 4 for the 73
galaxies in our sample. In each panel, the horizontal axis represents \xhii,
\xys, \xint, and \xold\ (from left to right), while the vertical axis shows
the percentage contributions at 5870\AA\ of stars in these four age bins. 
Stars younger than  $5\times10^{8}{\rm yr}$ dominate the emission in most
BCGs, but the galaxies also contain substantial fractions of older stars.

\subsection{Spectral fits}

\begin{figure*}
\centering
\includegraphics[angle=-90,width=\textwidth]{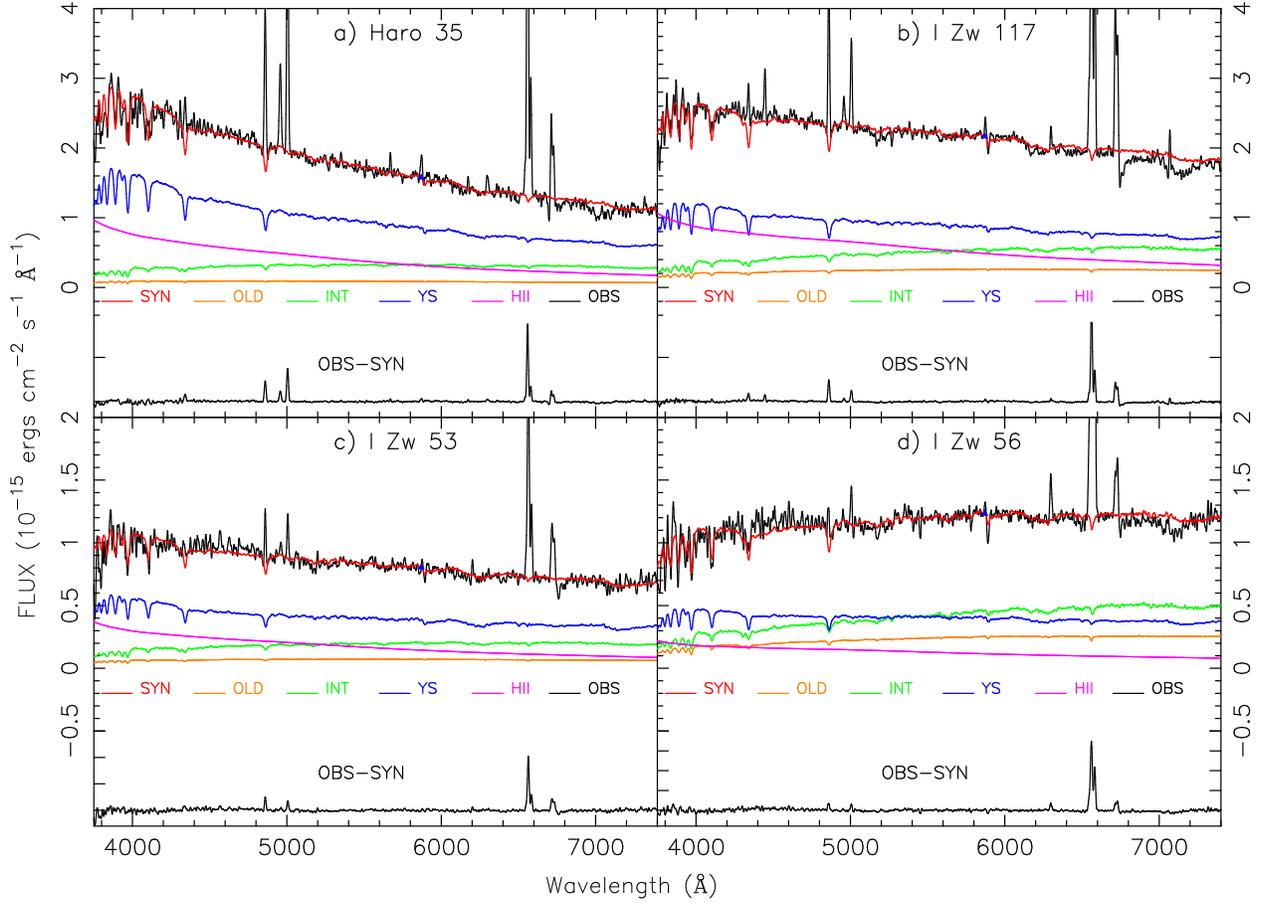}
\caption{
Comparison of synthetic spectra (red-solid lines) to the observed spectra of
four BCGs (corrected for Galactic reddening; black-solid lines): Haro 35, 
I Zw 117, I Zw 53, and I Zw 56. The contributions to the synthetic spectra
by old stars (OLD, $10^{10}{\rm yr}$), intermediate-age stars (INT, $10^{9}$, 
$5\times10^{9}{\rm yr}$), young stars (YS, $10^7$ -- $5\times10^{9}{\rm 
yr}$), and newly-born stars (\hii) are also shown. The emission line spectrum
appears in the OBS--SYN difference, at the bottom of each panel.}
\label{fig-spec}
\end{figure*}

Figure~\ref{fig-spec} illustrates the results of the spectral fits obtained
for four galaxies in our sample. Also shown are the contributions to the 
integrated spectrum by stars in the four age groups defined in section 4.3 
above (the relative contributions by the different stellar components to the 
total flux at $\lambda$5870 \AA\ are those listed in Tables 3 and 4 for these
galaxies).  Figure~\ref{fig-spec} shows that the synthetic spectra ({\bf SYN})
inferred from our population synthesis analysis provide good fits to the
observed spectra of BCGs ({\bf OBS}). The absorption wings of \hb, \hg\ and 
\hd\ in the observed spectra are also well reproduced by the models (the 
synthetic spectra do not include nebular emission lines). We find that, for 
some strong star-forming galaxies, the synthetic spectra do not provide very
good fits to the observed continuum spectra at wavelengths between 4300 and 
4800 \AA. This may arise from the presence of Wolf-Rayet (WR) features, such
as \niii\, features at $\lambda$4511 -- 4535, \nii$\lambda$4565, 
\nv$\lambda$4605, 4620, and \heii$\lambda$4686, in the observed spectra. We
plan to investigate this small discrepancy using evolutionary population 
synthesis models in a future paper.

\subsection{Evolutionary diagram}

\begin{figure}
\centering
\includegraphics*[angle=-90,width=\textwidth]{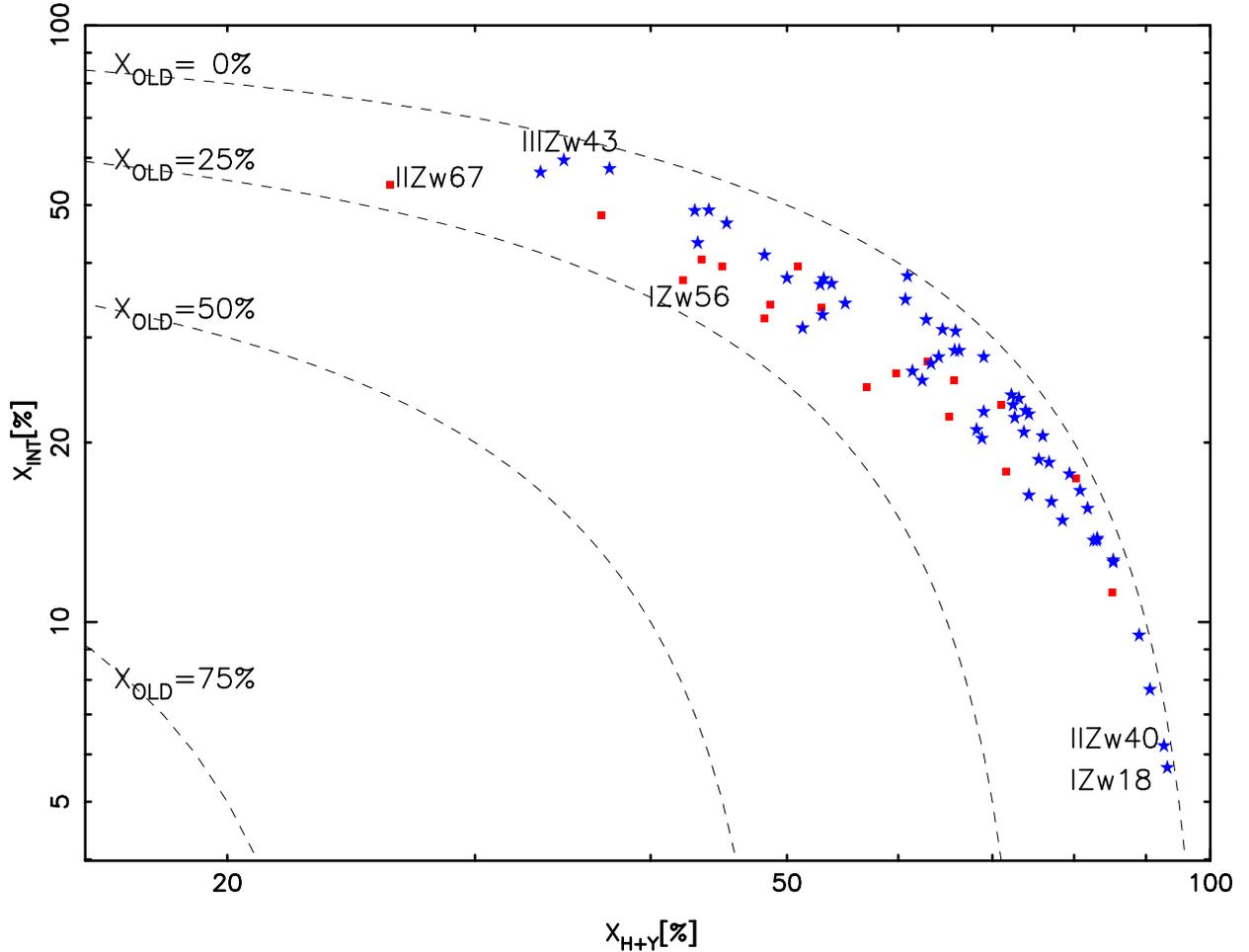}
\caption{
Results of the empirical population synthesis analysis for the galaxies 
with asymptotic (young-star) metallicities [Z/Z$_\odot$]=--0.5 (stars) 
and 0.0 (squares), condensed on an evolutionary diagram. 
The horizontal axis \xhy=\xhii+\xys\ is the fraction of light at 
$\lambda$5870 \AA\ due to star clusters between the $10^7$ and 
$5\times10^8{\rm yr}$ age bins and \hii\ regions, while the fraction 
\xint\ due to stars in the $10^9$ and $5\times10^9{\rm yr}$ age bins is 
plotted along the vertical axis. 
Dotted lines indicate lines of constant \xold\ ($10^{10}{\rm yr}$), as 
labeled.}
\label{fig-ed}
\end{figure}

The range of star formation histories inferred for the BGCs in our sample 
may be interpreted in terms of an evolutionary sequence. Following Cid
Fernandes \etal (2001b), we represent graphically the histories of star
formation of the galaxies in Figure \ref{fig-ed}, in a plane with abscissa
\xhy=\xhii+\xys\ and ordinate \xint.  Also shown as dashed lines in the 
figure are lines of constant \xold\ (\xhy+\xint+\xold=1).

The main result from Figure~\ref{fig-ed} is that the 73 BCGs in our 
sample define a sequence in \xhy\ and \xint\ over a relatively small
range of \xold. The range in \xold\ is slightly smaller for
galaxies with asymptotic (young-star) metallicity [Z/Z$_\odot$]=--0.5
(stars) than for those with asymptotic metallicity [Z/Z$_\odot$]= 0.0 
(squares). Some galaxies with `extreme' stellar populations are labelled
in Fig.~\ref{fig-ed}. In I Zw 56 and II Zw 67, for example, old stars
contribute up to $\sim 20$\% of the integrated flux at $\lambda$5870 
(see section 4.3 above). The galaxy III Zw 43 is that where 
intermediate-age stars contribute the most to the optical continuum
emission. In contrast, some galaxies, such as I Zw 18, II Zw 40, appear
to be almost `pure starbursts'. The small percentage of old and
intermediate-age stars found here for these galaxies is consistent
with the results from other recent studies, e.g., Smoker \etal\ (1999),
Aloisi \etal\ (1999) (see section 5.2 below).

Since the location of an individual galaxy in Figure~\ref{fig-ed} reflects
the evolutionary state of its stellar population, we can interpret the 
{\em distribution} of galaxies in this diagram as an {\em evolutionary} 
sequence. In particular, starbursting galaxies whose spectra are entirely
dominated by young massive stars (bottom right part of the diagram) will
presumably evolve toward larger \xint/\xhy values (top left part of the 
diagram) over a timescale of a few $10^{9}{\rm yr}$, once their bursts 
have ceased. This interpretation is supported by several observational 
facts. First, metal absorption-line features become stronger and galaxy 
colors become progressively redder as one moves from large \xhy\ to large
\xint\ along the sequence. Second, galaxies with large \xold\ (such as 
I Zw 56, II Zw 67) have spectra typical of a ``post-starburst'' galaxies,
with pronounced Balmer absorption lines characteristic of A-type stars and
with no strong emission lines. Finally, most BCGs in which WR features 
have been detected are located in the large-\xhy\ region of the diagram,
consistent with the young burst ages implied by the presence of WR stars
(\cite{schaerer99}). 

\subsection{Stellar Balmer absorption}

Accurate measurements of the H-Balmer emission lines are crucial 
to constrain the attenuation by dust, the star formation rate, 
the gas-phase abundances of chemical elements and the excitation 
parameter in galaxies (e.g., \cite{rosa02}). To measure with 
accuracy the Balmer emission-line fluxes of BCGs, we must account
for the contamination by underlying stellar absorption.

Different approaches have been used to correct Balmer emission-line
fluxes for underlying stellar absorption in BCGs. The simplest 
approach consists in adopting a constant equivalent width (1.5--2 \AA)
for all the Balmer absorption lines (e.g., \cite{skillman93}; 
\cite{popescu00}). Another standard correction consists in
determining the absorption equivalent width through an iterative
procedure, by assuming that the equivalent width is the same for all
Balmer lines and by requiring that the color excesses derived from 
\ha/\hb, \hb/\hg, and \hb/\hd\ be consistent (e.g., \cite{olive01}; 
\cite{cairos02}). In reality, however, the absorption equivalent 
width may not be the same for all H-Balmer lines.

\begin{figure}
\centering
\includegraphics*[angle=-90,width=\textwidth]{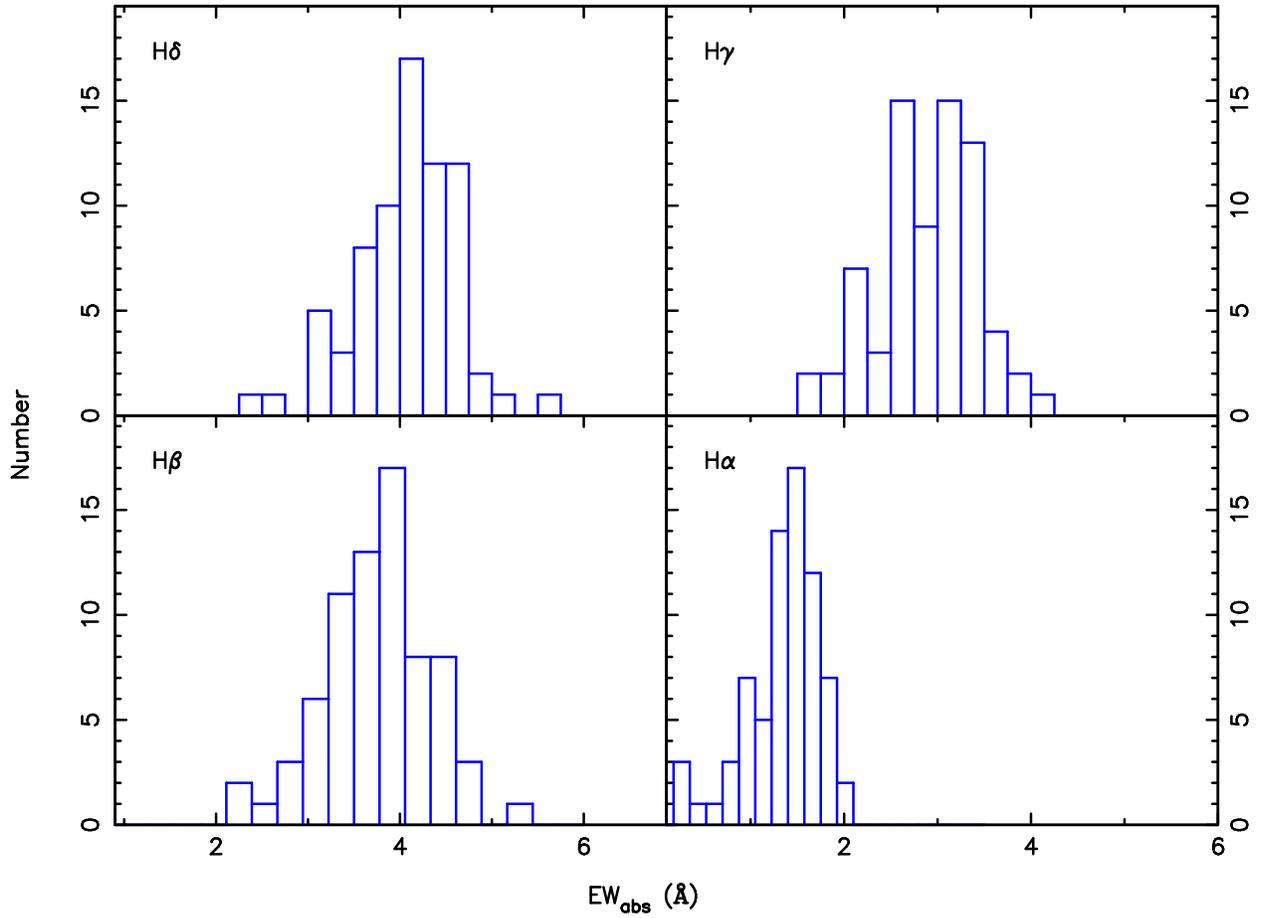}
\caption{
Distribution of the equivalent widths (in units of \AA) for the 
underlying Balmer absorption features, which were measured from 
the synthetic stellar population spectrum.}
\label{fig-abs}
\end{figure}

The advantage of the empirical population synthesis method used
above to fit the observed spectra of BCGs is that it provides
simultaneous fits to the continuum and stellar absorption features
of the galaxies. We have measured the absorption equivalent widths
of \ha, \hb, \hg, and \hd\ in the synthetic spectra fitted to all 
73 galaxies in our sample (last four columns of Table 3 and 4). 
Figure \ref{fig-abs} shows the distributions of the equivalent 
widths of the four lines. The equivalent widths of \hg, \hd, and 
\hb\ range typically between 2 \AA\ and 5 \AA\ (the distributions 
have different shapes for different lines), while that of \ha\ is 
typically less than 2 \AA. Hence, Figure \ref{fig-abs} indicates
that adopting a constant absorption equivalent width for all H-Balmer
lines is only a crude approximation. Our results are consistent with
those of Mas-Hesse \& Kunth (1999) and Olofsson (1995). 

We can correct with accuracy, therefore, the fluxes of Balmer emission 
lines for underlying stellar absorption in the spectra of BCGs. In
a forthcoming study, we will exploit the nebular emission-line 
spectra (OBS--SYN in Figure~\ref{fig-spec}) of BCGs to constrain 
the rates of star formation and gas-phase chemical element abundances
in these galaxies.

\section{Discussion}

\subsection{Age-metallicity degeneracy}

It is worth noting that the reason why we are able to constrain
simultaneously the ages and metallicities of the different 
stellar components in BCGs is that our approach is based on 
fitting stellar absorption lines, in addition to continuum 
fluxes. Age and metallicity have similar effects on the
continuum spectra of galaxies. However, some stellar absorption
features such as \hb, \hd, \hg\ and the G-band have been shown
to depend mostly on age, while others, such as \mgii\ and 
Fe$\lambda$5709, have been shown to depend mostly on 
metallicity (e.g., \cite{worthey97}; \cite{vazdekis99}; 
\cite{kong01}).

In this paper, we use the \caii\ K $\lambda$3933, \hd\ 
$\lambda$4102, CN $\lambda$4200, G band $\lambda$4301, \hg\ 
$\lambda$4340 and \mgii\ $\lambda$5176 absorption features to 
determine the stellar population of BCGs. These include both
age-sensitive and metal-sensitive features. Some of these 
absorption features may not be available in some spectra, 
which may lack, for example, \hd\ and \hg. Also, some absorption 
features may be very shallow, and hence difficult to measure 
with high accuracy. For most galaxies, however, we find that the
results of the empirical population synthesis analysis provide 
coarse but useful constraints on the histories of star formation
and metal enrichment.

\subsection{Population synthesis versus CMD analysis}

The most straightforward way to observe the star formation history 
of galaxies is through deep single-star photometry. This allows one
to directly identify stars in various evolutionary phases through 
their positions in a color-magnitude diagram (CMD) containing 
the fossil record of the star formation history. During the past few
years, intense activity has focused on the reconstruction of the 
star formation histories of nearby galaxies using this approach. 
Several of these studies were performed on BCGs and led to new
constraints on the ages of the oldest stars present in these galaxies
(\cite{aloisi99}; \cite{crone02}; \cite{papaderos02}; \cite{oestlin98}). 
The general conclusions from these studies are that BCGs contain
evolved stellar populations, and that their star formation histories 
have been discontinuous.  The basic shortcoming of this approach is 
that it cannot be used in galaxies at large distances. In addition,
the CMD-analysis method is also subject to several uncertainties 
such as distance determination, extinction, and contamination of 
stellar colors by gaseous emission.
 
Our results of the EPS analysis of 73 BCGs are consistent with those
of the CMD analysis of a few nearby galaxies, supporting the finding
that BCGs are old galaxies with intermittent star formation. The 
evolutionary diagram constructed in Figure~\ref{fig-ed} above
shows that the EPS method is not only able to recognize composite
systems from a handful of observable absorption-line features and 
continuum fluxes, but also to provide a rough description of the 
evolutionary state of the stellar components. Therefore, the EPS 
method provides a convenient tool for the study of stellar components
and star formation histories of galaxies. It can be easily applied 
to the spectra gathered by large spectroscopic galaxy surveys.

\section{Summary}

We have presented the results of an empirical population synthesis 
study of a sample of 73 blue compact galaxies. Our main goal was to
study the stellar components of BCGs. We have constrained the star
formation histories of BCGs by comparing observed stellar absorption 
features and continuum fluxes with a library of star cluster spectra.
Our conclusions can be summarized as follows:

BCGs present a variety of star formation histories, as inferred from
the wide spread in stellar absorption equivalent widths and continuum
colors from galaxy to galaxy. BCGs are typically age-composite stellar
systems, in which different stellar components are clearly 
distinguished: the current starburst, an underlying older population,
and some intermediate-age population.

A quantitative analysis indicates that the nuclei of some BCGs 
are dominated by young components and the star-forming process is 
still ongoing. In most of BCGs, stars older than 1~Gyr contribute
significantly to the integrated optical emission. The contribution
by these stars can exceed 40\% in some cases. Overall, the stellar
populations of BCGs suggest that they are old galaxies undergoing
intermittent star formation episodes; a typical BCGs is not presently
forming its first generation of stars. We also find that the 
attenuation by dust is typically very small in the BCGs in our
sample.

Our results are consistent with the results from deep imaging 
observations using HST and large ground-based telescopes. The 
virtue of the EPS approach is that it is applied to integrated 
galaxy spectra. This method should be useful, therefore, for
interpreting the spectra garthered by large spectroscopic galaxy
surveys.

The EPS approach also provides accurate spectral fits of observed
galaxy spectra. From these spectral fits, it is possible to measure
with accuracy, in particular, the absorption strengths of stellar
Balmer lines and to correct the observed Balmer emission-line fluxes
for underlying stellar absorption. The pure emission-line spectra
of the BCGs in our sample, resulting from the subtraction of the 
synthetic spectral fits from the observed spectra, will be presented
in a forthcoming paper. 

\begin{acknowledgements}

We thank an anonymous referee for helpful comments and constructive 
suggestions, which helped us improve the paper. We are very grateful 
to Dr. R. Cid Fernandes for giving us the EPS computer program, and 
kindly assisting us in its use. This work is based on observations 
made with the 2.16m telescope of the Beijing Astronomical 
Observatory(BAO) and supported by the Chinese National Natural 
Science Foundation (CNNSF 10073009).  S.C. thanks the Alexander
von Humboldt Foundation, the Federal Ministry of Education and Research,
and the Programme for Investment in the Future (ZIP) of the German 
Government for support.  X.K. has been financed by the Special Funds 
for Major State Basic Research Projects of China and the Alexander 
von Humboldt Foundation of Germany.
\end{acknowledgements}

\end{document}